\documentclass[a4paper,12pt]{article}
\usepackage{cite}
\usepackage{amsmath}
\usepackage{amssymb}
\newcommand {\oks}[2]{{\raise0.7ex\hbox{${\scriptstyle #1}$}\!\mathord{\left/
{\vphantom{{1}{2}}}\right.\kern-\nulldelimiterspace}\!\lower0.7ex
\hbox{${\scriptstyle #2}$}}}
\begin{document}

\title{Spin light of neutrino in matter \\ and
electromagnetic fields}

\author{A.Lobanov, A.Studenikin\thanks{E-mail: studenik@srd.sinp.msu.ru}}
\date{}
\maketitle

\begin{center}
{\em Department of Theoretical Physics, Moscow State University,
119992, Moscow, Russia}
\end{center}
%\rightline DTP-MSU/00-08

\begin{abstract}
A new type of electromagnetic radiation by a neutrino with
non-zero magnetic (and/or electric) moment moving in background
matter and electromagnetic field is considered. This radiation
originates from the quantum spin flip transitions and we have
named it as "spin light of neutrino"($SL\nu$). The neutrino
initially unpolarized beam (equal mixture of $\nu_{L}$ and
$\nu_{R}$) can be converted to the totally polarized beam composed
of only $\nu_{R}$ by the neutrino spin light in matter and
electromagnetic fields. The quasi-classical theory of this
radiation is developed on the basis of the generalized
Bargmann-Michel-Telegdi equation. The considered radiation is
important for environments with high effective densities, $n$,
because the total radiation power is proportional to $n^{4}$. The
spin light of neutrino, in contrast to the Cherenkov or transition
radiation of neutrino in matter, does not vanish in the case of
the refractive index of matter is equal to unit.  The specific
features of this new radiation are: (i) the total power of the
radiation is proportional to $\gamma ^{4}$, and (ii) the radiation
is beamed within a small angle $\delta \theta \sim \gamma^{-1}$,
where $\gamma$ is the neutrino Lorentz factor. Applications of
this new type of neutrino radiation to astrophysics, in particular
to gamma-ray bursts, and the early universe should be important.
\end{abstract}

 There exist at present convincing evidences
in favour of neutrino non-zero mass and mixing, obtained in the
solar and atmospheric experiments (see \cite{Bil02} for a review
on the status of neutrino oscillations). Apart from masses and
mixing non-trivial neutrino electromagnetic properties such as
non-vanishing magnetic, $\mu$, and electric, $\epsilon$, dipole
moments are carrying features of new physics.  It is  believed
that non-zero neutrino magnetic moment could have an important
impact on astrophysics and cosmology.

It is well known \cite{LSFS7780} that in the minimally extended
Standard Model with $SU(2)$-singlet right-handed neutrino the
one-loop radiative correction generates neutrino magnetic moment
which is proportional to neutrino mass
\begin{equation}
\mu_{\nu}={3 \over {8 \sqrt{2}\pi^{2}}}eG_{F}m_{\nu}=3 \times
10^{- 19}\mu_{0}\bigg({m_{\nu} \over {1 \mathrm{eV}}}\bigg),
\end{equation}
where $\mu_{0}=e/2m$ is the Bohr magneton, $m_{\nu}$ and $m$ are
the neutrino and electron masses. There are also models
\cite{KBMVVFY76-87} in which much large values for magnetic
moments of neutrinos are predicted. So far, the most stringent
laboratory constraints on the electron neutrino magnetic moment
come from elastic neutrino-electron scattering experiments:
$\mu_{\nu_{e}} \leqslant 1.5 \times 10^{-10} \mu_{0}$ \cite{PDG}.
More stringent constraints are obtained from astrophysical
considerations \cite{Raf96}.

In this paper we study a new mechanism for emission of photon by
the massive neutrino in presence of matter, assuming that the
neutrino has an intrinsic magnetic (and/or electric) dipole
moment. This phenomenon can be expressed as a process
\begin{equation}\label{b}
\nu \rightarrow \nu + \gamma
\end{equation}
that is the transition from a flavour neutrino in the initial
state to the same flavour neutrino plus a photon in the final
state. The mechanism under consideration can be also effective in
the case of neutrino transitions with change of flavour if
neutrino transition moment is not zero.

Different other processes characterized by the same signature
of Eq.(\ref{b}) have been considered  previously:

 i) the photon radiation by massless neutrino
$(\nu_{i} \rightarrow \nu_{j} + \gamma,\; i=j)$ due to the vacuum
polarization loop diagram  in presence of an external magnetic
field \cite{GalNik72,Sko76};

ii) the photon radiation by massive neutrino with non-vanishing
magnetic moment in constant magnetic and electromagnetic wave
fields \cite{BorZhuTer88,Sko91};

iii) the Cherenkov radiation due to the non-vanishing neutrino
magnetic moment in homogeneous and infinitely extended medium
which is only possible if the speed of neutrino is larger than the
speed of light in medium \cite{Rad75,GriNeu93};

iv) the transition radiation due to non-vanishing neutrino
magnetic moment which would be produced when the neutrino crosses
the interface of two media with different refractive indices
\cite{SakKur94-95,GriNeu95};

v) the Cherenkov radiation by  massless neutrino due to its
induced charge in medium \cite{OliNiePal96};

vi) the Cherenkov radiation by massive and massless  neutrino in
magnetized medium \cite{MohSam96, IoaRaf97};

vii) the neutrino radiative decay $(\nu_{i} \rightarrow \nu_{j} +
\gamma, \; i\not=j)$ in external fields and media or in vacuum
\cite{GvoMikVas92,Sko95,ZhuEmiGri96,KacWun97,TerEmi03}.

The process we are studying in this paper has never been
considered before. We discover a mechanism for electromagnetic
radiation generated by the neutrino magnetic (and/or electric)
moment rotation which occurs due to electroweak interaction with
the background environment. It should be noted that generalization
to the case of a photon emission by neutrino due to the neutrino
transition magnetic moment is straightforward. If neutrino is
moving in matter and an external electromagnetic field is also
superimposed, the total power of this radiation contains three
terms which originate from (i) neutrino interaction with particles
of matter, (ii) neutrino interactions with electromagnetic field,
(iii) interference of the mentioned above two types of
interactions. This radiation can be named as "spin light of
neutrino" ($SL\nu$) in matter and electromagnetic field to
manifest the correspondence with the magnetic moment dependent
term in the radiation of an electron moving in a magnetic field. A
review on the spin light of electron can be found in
\cite{BorTerBag95}. Whereas the radiation of a neutral particle
moving in external electromagnetic field in the absence of matter
has been considered previously starting from \cite{TerBagKha65},
the mechanism of radiation produced by interaction with matter is
considered in this our paper for the first time. It should be
emphasize that the neutrino spin light can not be described as the
Cherenkov radiation.

The $SL\nu$ in the background matter ( similar to the radiation by
neutrino moving in the magnetic field \cite{BorZhuTer88})
originates from the quantum spin flip transitions $\nu_{L}
\rightarrow \nu_{R}$. Within the quantum approach the
corresponding Feynman diagram of the proposed new process is the
standard one-photon emission diagram with the initial and final
neutrino states described by the "broad lines" that account for
the neutrino interaction with matter (given, for instance, by the
effective Lagrangian of Eq.(\ref{Lag}) below). In this paper we
develop the quasi-classical approach to the radiation process when
the neutrino recoil can be neglected. This approach is valid for a
wide range of neutrino and photon energies that could be of
particular interest for different astrophysical and cosmological
applications. For example, if the initial neutrino energy is about
$10 \ \ MeV$ the allowed range of the radiated photon energies
span up to gamma-rays.

We should like also to emphasize here that the initially
unpolarized neutrino beam (equal mixture of active left-handed and
sterile right-handed neutrinos) can be converted to the totally
polarized beam composed of only $\nu_{R}$ due to the spin light in
contrast to the Cherenkov radiation which can not produce the
neutrino spin self-polarization effect.

Our approach is based on the quasi-classical Bargmann-Michel-Telegdi (BMT)
equation \cite{BMT59}
\begin{equation}\label{1}
{dS^{\mu} \over d\tau} =2\mu
 \Big\{ F^{\mu\nu}S_{\nu} -u^{\mu}(
u_{\nu}F^{\nu\lambda}S_{\lambda} ) \Big\} +2\epsilon \Big\{
{\tilde
F}^{\mu\nu}S_{\nu} -u^{\mu}(u_{\nu}{\tilde
F}^{\nu\lambda}S_{\lambda}) \Big\},
\end{equation}
that describes evolution of the spin $S_{\mu}$ of a neutral
particle with non-vanishing magnetic, $\mu$, and electric,
$\epsilon$, dipole moments in electromagnetic field, given by its
tensor $F_{\mu \nu}$ {\footnote{Within this approach neutrino spin
relaxation in stochastic electromagnetic fields without account
for matter effects was considered in \cite{LoeSto89}.}}. This form
of the BMT equation corresponds to the case of the particle moving
with constant speed, $\vec \beta={\mathrm{const}}$,
$u^\mu=(\gamma,\gamma \vec \beta)$, in presence of an
electromagnetic field $F_{\mu\nu}$. The spin vector satisfies the
usual conditions, $S^2=-1$ and $S^{\mu}u_\mu =0$. Note that the
term proportional to $\epsilon$ violates $T$ invariance.

In our previous studies \cite{EgoLobStu00,LobStu01} (see also
\cite{Stu_Mor_02}) we have shown that  the Lorentz
invariant generalization of Eq.(\ref{1}) for the case when effects
of neutrino weak interactions are taken into account can be
obtained by the following substitution of the electromagnetic
field tensor $F_{\mu\nu}=(\vec E,\vec B)$:
\begin{equation}\label{2}
F_{\mu\nu} \rightarrow E_{\mu\nu}= F_{\mu\nu}+G_{\mu\nu},
\end{equation}
where the tensor $G_{\mu\nu}$ accounts for the neutrino
interactions with particles of the environment. The derivation of
the quasi-classical Lorentz invariant neutrino spin evolution
equation taking into account general types of neutrino
non-derivative interactions with external fields is given in
\cite{DvoStu02}. Within the quantum approach the neutrino spin
flip under the influence of different types of interactions was
also considered in \cite{BergGrosNar99}.

In evaluation of the tensor $G_{\mu\nu}$ we demand that the
neutrino evolution equation must be linear  over the neutrino
spin, electromagnetic field and such characteristics of matter
(which is composed of different fermions, $f=e,\ n,\ p...$) as
fermions currents
\begin{equation}
j_{f}^\mu=(n_f,n_f\vec v_f),
\label{3}
\end{equation}
and fermions polarizations
\begin{equation}
\lambda^{\mu}_f =\Bigg(n_f (\vec \zeta_f \vec v_f ),
n_f \vec \zeta _f \sqrt{1-v_f^2}+
{{n_f \vec v_f (\vec \zeta_f \vec v_f )} \over {1+\sqrt{1-
v_f^2}}}\Bigg).
\label{4}
\end{equation}
Here $n_f$, $\vec v_f$, and $\vec \zeta_f \ (0\leqslant |\vec
\zeta_f |^2 \leqslant 1)$ denote, respectively, the number
densities of the background fermions $f$, the speeds of the
reference frames in which the mean momenta of fermions $f$ are
zero, and the mean values of the polarization vectors of the
background fermions $f$ in the above mentioned reference frames.
The mean value of the background fermion $f$ polarization vector,
$\vec \zeta_f$, is determined by the two-step averaging of the
fermion relativistic spin operator over the fermion quantum state
in a given electromagnetic field and  over the fermion statistical
distribution density function. Thus, in general case of neutrino
interaction with different background fermions $f$ we introduce
for description of matter effects antisymmetric tensor
\begin{equation}
G^{\mu \nu}= \epsilon ^{\mu \nu \rho \lambda}
g^{(1)}_{\rho}u_{\lambda}- (g^{(2)\mu}u^\nu-u^\mu g^{(2)\nu}),
\label{7}
\end{equation}
where
\begin{equation}
g^{(1)\mu}=\sum_{f}^{} \rho ^{(1)}_f j_{f}^\mu
+\rho ^{(2)}_f \lambda _{f}^{\mu}, \ \
g^{(2)\mu}=\sum_{f}^{} \xi ^{(1)}_f j_{f}^\mu
+\xi ^{(2)}_f \lambda _{f}^{\mu},
\label{8}
\end{equation}
(summation is performed over the fermions $f$ of the background).
The explicit
expressions for the coefficients $\rho_{f}^{(1),(2)}$ and
$\xi_{f}^{(1),(2)}$
could be found if the particular
model of neutrino interaction is chosen.
For example, if one consider the electron neutrino propagation in moving and
polarized gas of electrons within the extended standard
model supplied with $SU(2)$-singlet right-handed neutrino
$\nu_{R}$, then the neutrino effective
interaction Lagrangian reads
\begin{equation}\label{Lag}
L_{eff}=-f^\mu \Big(\bar \nu \gamma_\mu {1+\gamma^5 \over 2} \nu
\Big),
\end{equation}
where
\begin{equation}
f^\mu={G_F \over \sqrt2}\Big((1+4\sin^2 \theta _W) j^\mu_e -
\lambda ^\mu _e\Big).
\end{equation}
In this case the coefficients $\rho_{e}^{(1),(2)}$  are
\begin{equation}\label{rho}
\rho^{(1)}_e={\tilde{G}_F \over {2\sqrt{2}\mu }}\,, \qquad
\rho^{(2)}_e =-{G_F \over {2\sqrt{2}\mu}}\,,
\end{equation}
where $\tilde{G}_{F}={G}_{F}(1+4\sin^2 \theta _W).$

 In the usual notations the antisymmetric tensor $G_{\mu
\nu}$ can be written in the form,
\begin{equation}
G_{\mu \nu}= \big(-\vec P,\ \vec M),
\label{9}
\end{equation}
where
\begin{equation}
\vec M= \gamma \big\{(g^{(1)}_0 \vec \beta-\vec g^{(1)})
- [\vec \beta \times \vec g^{(2)}]\big\}, \
\vec P=- \gamma \big\{(g^{(2)}_0 \vec \beta-\vec g^{(2)})
+ [\vec \beta \times \vec g^{(1)}]\big\}.
\label{10}
\end{equation}
It worth to note that the substitution (\ref{2}) implies that the
magnetic $\vec B$ and electric $\vec E$ fields are shifted by the
vectors $\vec M$ and $\vec P$, respectively:
\begin{equation}
\vec B \rightarrow \vec B +\vec M, \ \ \vec E \rightarrow \vec E -
\vec P.
\label{11}
\end{equation}

We should like to emphasize here that  precession of the neutrino
spin can originate not only due to neutrino magnetic moment
interaction with external electromagnetic fields but also due to
the neutrino weak interaction with particles of the background
matter. This is a very important point for understanding of the
nature of the neutrino spin light in non-magnetized matter. In
order to demonstrate how the neutrino spin procession appears in
the background matter we consider below the neutrino spin
evolution in matter in the absence of electromagnetic fields. We
start with Eq.(\ref{1}) and for simplicity neglect the neutrino
electric dipole moment, $\epsilon=0$. Then, in the absence of
external electromagnetic field, we get the neutrino spin evolution
equation in non-moving matter:
\begin{equation}
{dS^{\mu} \over d\tau} =2\mu
 \Big\{ G^{\mu\nu}S_{\nu} -u^{\mu}(
u_{\nu}G^{\nu\lambda}S_{\lambda} ) \Big\}\label{s_m}.
\end{equation}
The tensor $G_{\mu\nu}$ is given by Eqs.(\ref{7}),(\ref{8}) and
(\ref{rho}).  For the further simplifications we consider
unpolarized ($\lambda _{f}^{\mu}=0$) matter composed of only one
type of fermions, so that there is now summation over $f$ in
definition of $g^{(1)\mu}$ and we shall omit the index $f$. Then
we get
\begin{equation}
G^{\mu\nu}=\gamma \rho^{(1)}n \begin{pmatrix}{0}&{0}& {0}&{0} \\
{0}& {0}& {-\beta_{3}}&{\beta_{2}} \\
{0}&{\beta_{3}}& {0}&{-\beta_{1}} \\
{0}&{-\beta_{2}}& {\beta_{1}}& {0} \end{pmatrix}, \label{g}
\end{equation}
where $\vec \beta = (\beta_{1}, \beta_{2}, \beta_{3})$ is the
neutrino three-dimensional speed. It is easy to show that
\begin{equation}
u_{\nu}G^{\nu\mu}=0\label{uG},
\end{equation}
and from (\ref{s_m}) we get the equation for the neutrino spin
evolution in unpolarized and non-moving matter:
\begin{equation}
{dS^{\mu} \over d\tau} =2\mu G^{\mu\nu}S_{\nu}.
\end{equation}
In the laboratory reference frame the corresponding equation for
the three-dimensional neutrino spin is
\begin{equation}
{d{\vec S} \over dt} = 2\mu\rho^{(1)}n[\vec S \times \vec \beta ].
\end{equation}
If neutrino is propagating along the OZ axis, $\vec \beta= (0,0,
\beta)$, then solutions of these equations for the neutrino spin
components are given by
\begin{equation}
S^{1}=S_{0}^{\perp} \cos \omega t,\, \,
 S^{2}=S_{0}^{\perp} \sin \omega t, \,\,
 S^{3}=S^{3}_{0}, S^{0}=S^{0}_{0},
 \end{equation}
where
\begin{equation}
\omega= 2\mu\rho^{(1)}n\beta,
\end{equation}
$S_{0}^{\perp}$ and $S^{3,0}_{0}$ are constants determined by the
initial conditions.

From the above consideration it follows that if the initial
neutrino state is not polarized longitudinally in respect to the
neutrino momentum then the neutrino spin precession in the
background matter always occurs. In its turn, a neutrino with
processing magnetic momentum have to emit electromagnetic
radiation. This is just the radiation which we have called the
spin light of neutrino.

Recently we have derived the total radiation power of a neutral
unpolarized fermion with anomalous magnetic moment \cite{LobPav00}
{\footnote {The generalization to the case of neutrino with
non-zero electric dipole moment in electromagnetic field can be
found in \cite{LobPav_Vest00}}}. In that derivation we have
supposed that the spin dynamics of  a neutral particle is governed
by the Bargmann-Michel-Telegdi equation and that the energy of the
radiated photons is much less than the particle energy. Now in
order to treat the electromagnetic radiation by the neutrino
moving in background matter and electromagnetic field we use the
substitution prescription of Eq.(\ref{2}) and for the total
radiation power get (for simplicity the case of $T$-invariant
model of neutrino interaction is considered hereafter)
\begin{equation}\label{power} I={16 \over
3}\mu^2 \Big[4({\mu^2}{u_\rho}{\tilde E^{\rho \lambda}} {\tilde
E_{\lambda\sigma}}{u^\sigma})^2+ {\mu^2}{u_\rho}{{\dot { \tilde
E}}^{\rho\lambda}}{\dot {\tilde E}_{\lambda \sigma}} {u^\sigma}
\Big].
\end{equation}
We also get for the solid angle distribution of the radiation
power :
\begin{equation}\label{ra11}
\begin{array}{c}
\!\!\!\displaystyle\frac{dI}{d\Omega}=\displaystyle\frac{\mu^2}{\pi
\gamma (ul)^5} \Big\{\big[ 4({\mu^2}{u_\rho}{\tilde E
^{\rho\lambda}}{\tilde E_{\lambda\sigma}}{u^\sigma})^2+
({\mu^2}{u_\rho}{{\dot {\tilde E} }^{\rho \lambda}}{{\dot {\tilde
E} }_{\lambda \sigma}}
{u^\sigma})\big] (ul)^2 +\\[14pt]\quad
+4({\mu^2}{u_\rho}{\tilde E ^{\rho\lambda}}{\tilde E
_{\lambda\sigma}} {u^\sigma}) ({\mu}{u_\rho}{\tilde E
^{\rho\lambda}}l_{\lambda})^2+
({\mu}{u_\rho}{{\dot {\tilde E }}^{\rho\lambda}}l_{\lambda})^2 +\\[14pt]
\qquad+4\mu^2({\mu}{u_{\rho}}{\tilde E ^{\rho\lambda}}l_{\lambda})
e^{\mu\nu\rho\lambda} u_{\mu}{\dot {\tilde E
}}_{\nu\sigma}{u^{\sigma}} {\tilde E _{\rho\delta}}{u^{\delta}}
{l_{\lambda}}\Big\},
\end{array}
\end{equation}
here $\tilde E_{\mu\nu}=-\oks{1}{2} \epsilon
_{\mu\nu\alpha\beta}E^{\alpha\beta}$ is the dual tensor
$E_{\mu\nu},\,$ $l^{\mu}=(1, \vec l)$ and   $\vec l$ is the unit
vector pointing the direction of radiation.  The derivatives in
the right-hand side of Eqs.(\ref{power}) and (\ref{ra11}) are
taken with respect of proper time $\tau$ in the rest frame of the
neutrino.

Using Eqs. (\ref{11}) and (\ref{power}) we find the total
radiation power as a function of the magnetic field strength in
the rest frame of the neutrino, $\vec B_{0}$, and the vector $\vec
M_{0}$ which accounts for effects of  the neutrino interaction
with moving and polarized matter:
\begin{equation}\label{0ra13}
I=\frac{16}{3}\mu^4 \Big[\big(2{\mu}(\vec B_{0}+\vec
M_{0}\big)^{2}\big)^2 +(\dot{\vec {B}_{0}} +\dot{\vec
{M}_{0}})^{2}\Big],
\end{equation}
where
\begin{equation}
\vec B_0=\gamma\Big(\vec
B_{\perp}
+{1 \over \gamma} \vec B_{\parallel} + \sqrt{1-\gamma^{-2}}
\Big[{\vec E_{\perp} \times \vec n}\Big]\Big),
\end{equation}

\begin{equation}
\vec {M_0}=\vec {M}{_{0_{\parallel}}}+\vec {M_{0_{\perp}}},
\label{M_0}
\end{equation}
\begin{equation}
\begin{array}{c}
\displaystyle \vec {M}_{0_{\parallel}}=\gamma\vec\beta{n_{0} \over
\sqrt {1- v_{e}^{2}}}\left\{ \rho^{(1)}\left(1-{{\vec v}_e
\vec\beta \over {1- {\gamma^{-2}}}} \right)\right.
-\\- \displaystyle\rho^{(2)}\left.
\left(\vec\zeta_{e}\vec\beta \sqrt{1-v^2_e}+ {(\vec \zeta_{e}{\vec
v}_e)(\vec\beta{\vec v}_e) \over 1+\sqrt{1-v^2_e} }\right){1 \over
{1- {\gamma^{-2}}}} \right\}, \label{M_0_parallel}
\end{array}
\end{equation}
\begin{equation}\label{M_0_perp}
\begin{array}{c}
\displaystyle \vec {M}_{0_{\perp}}=-\frac{n_{0}}{\sqrt {1-
v_{e}^{2}}}\Bigg\{ \vec{v}_{e_{\perp}}\Big(
\rho^{(1)}+\displaystyle\rho^{(2)}\frac
{(\vec{\zeta}_{e}{\vec{v}_e})}
{1+\sqrt{1-v^2_e}}\Big)+{\vec{\zeta}_{e_{\perp}}}\rho^{(2)}\sqrt{1-v^2_e}\Bigg\},
\end{array}
\end{equation}
and $n_0$ is the invariant number density of matter given in the
reference frame for which the total speed of matter is zero.

It is evident that the total radiation power, Eq.(\ref{0ra13}),
 is composed of the three contributions,
\begin{equation}
I=I_{F}+I_{G}+I_{FG},
\end{equation}
where $I_{F}$ is the radiation power due to the neutrino magnetic
moment interaction with the external electromagnetic field,
$I_{G}$ is the radiation power due to the neutrino weak
interaction with particles of the background matter, and $I_{FG}$
stands for the interference effect of electromagnetic and weak
interactions. It should be pointed out that  the electromagnetic
contribution $I_{F}$ to the radiation of a neutrino (or a neutral
fermion) in different field configurations has been considered in
literature (see, for example,
(\cite{TerBagKha65,BorZhuTer88,Sko91}), whereas the contribution
to radiation by neutrino moving in matter, $I_{G}$, and also the
interference term $I_{FG}$ have never been considered before.

These three types of neutrino radiation have common nature: they
originates as effects of the neutrino interactions with the
external electromagnetic field and background matter under which
the neutrino spin is rotating. As it has been pointed out above,
the discussed here radiation cannot be treated as the neutrino
Cherenkov \cite{Rad75,GriNeu93,OliNiePal96,MohSam96,IoaRaf97} and
transition \cite{SakKur94-95,GriNeu95} radiations. Contrary to the
Cherenkov and transition radiations the considered new type of
neutrino radiation is  not forbidden when the refractive index of
photons in the background environment is equal to
$n^{ref}_{\gamma}=1$. In order to highlight this distinction we
consider here the particular case of $n^{ref}_{\gamma}=1$, however
generalization to the case $n^{ref}_{\gamma}>1$ is
straightforward.

One of the most important properties of the $SL\nu$ in the
background matter is the strong dependence of the total radiation
power on the density of matter,
\begin{equation} \label{G}
I_{G} \sim {n_{0}^{4}}.
\end{equation}

The total radiation power, Eq.(\ref{0ra13}),  contains different
terms in respect to dependence on the neutrino magnetic moment
$\mu$. If we consider non-derivative terms and compare
contributions to the neutrino spin light radiation power from the
interaction with matter and electromagnetic field we come to the
conclusion that, as it follows from Eq.(\ref{rho}), the ratio
$I_{G}/I_{F}$ is proportional to $\mu^{-4}$. From Eq.(\ref{0ra13})
it is also obvious that the ratios of the radiation power in
matter with varying density (in varying electromagnetic fields) to
the radiation power in matter with constant density (in constant
electromagnetic fields) are proportional to $\mu^{2}$. Accounting
for smallness of the neutrino magnetic moment, it follows that the
proposed new mechanism of neutrino spin light radiation could be
efficient in the environments with varying densities of matter
(with varying electromagnetic fields).

We consider at first the $SL\nu$ in presence of constant density
matter and constant magnetic field. For definiteness we assume
that the spin light radiation is produced by the electron neutrino
$\nu_{e}$ moving in unpolarized ($\zeta_e=0$) matter composed of
only electrons. Then for the total spin light radiation power we
get
\begin{equation}\label{BG}
I={64 \over 3} \mu^{6} \gamma^{4}\Big[ \big( \vec B_{ \perp} + {1
\over \gamma} \vec B_{
\parallel} +  n\rho^{(1)} \vec \beta (1- \vec \beta \vec
v_e) - {1 \over \gamma} n \rho^{(1)} \vec v_{e_\perp}\big)^{2}
\Big]^{2},
\end{equation}
where the terms proportional to ${\gamma^{-2}}$ in the brackets
are neglected. Here we use the notation,
\begin{equation}
n={n_0 \over \sqrt{1-v_{e}^{2}}},
\end{equation}
$\vec B_{ \perp , \parallel}$ are the transversal and longitudinal
magnetic field components in respect to the neutrino motion, $\vec
v_{\perp}$ is the transversal component of the matter speed in the
laboratory frame of reference. From Eq.(\ref{BG}) it follows that
the correlation term $I_{FG}$ is suppressed  in respect to the
terms $I_F$ and $I_G$ by the presence of additional neutrino
Lorentz factor. That is why there is a reason to compare
contributions to the radiation power produced by the neutrino
interaction with matter which is moving longitudinally with
non-relativistic speed, $v \ll 1$,
\begin{equation}\label{I_G}
I_{G}={64 \over 3} \mu^{6}\gamma^{4}\big(  n\rho^{(1)} \beta
\big)^{4},
\end{equation}
and the contribution to the radiation power produced by the
interaction with the transversal magnetic field{\footnote{This our
result is smaller by a factor $1/2$ relative to the result of
\cite{TerBagKha65} for the radiation power of the polarized
neutral particle.}},
\begin{equation}\label{I_F}
I_{F}={64 \over 3} \mu^{6}\gamma^{4}B_{ \perp}^{4}.
\end{equation}
Note that from Eq.(\ref{I_G}) it follows that in the case of the
standard model, when the neutrino magnetic moment is proportional
to its mass, and constant neutrino speed the radiation power goes
with the neutrino mass squared, as one might expect. If we take
$B_{\perp} \sim 3\times 10^{5}\,\mathrm{G}$ and $n \sim 10^{23}\,
\mathrm{cm}^{-3}$ that corresponds to the case of the solar
convective zone and $\mu_{\nu_e} \sim 10^{-18} \mu_{0}$, then we
get that the matter term in the spin light exceeds the transversal
magnetic field term by a factor of $\sim 2 \times 10^{26}$. With
increasing of the neutrino magnetic moment due to the inverse
proportionality $\rho^{(1)} \sim \mu_{\nu_e}^{-1}$ the ratio $
I_{G}/ I_{F}$ decreases and becomes equal to unit only for
$\mu_{\nu_e} \sim 3 \times 10^{-12}\mu_{0}$.

Let us turn to consideration of the contribution to the $SL\nu$ that
is generated by the neutrino interaction with matter of varying
density. We again assume that matter is composed of electrons and
neutrino interaction with matter is given by Lagrangian of
Eq.(\ref{Lag}). Since $\dot n_e = \gamma \vec \beta \vec {\nabla}
n_e$, we get for the total power of the neutrino spin light in
this case
\begin{equation}\label{0ra}
I_{G}=\frac{2}{3}\mu^{2}\gamma^{4}
\left\{\frac{1}{2}{\tilde{G}_F^{4}n_e}^{4}+\tilde{G}_F^{2}
\left[({\vec{\beta}}{\vec{\nabla}})n_e\right]^{2} \right\}\!.
\end{equation}
If we consider the case of moving and polarized matter then the
effective number density depends on the value of the total mater
speed $\vec v_{e}$ and polarization $\vec \zeta _{e}$, as well as
on neutrino speed $\vec \beta$ and correlations between these
three vectors (see \cite{LobStu01,GriLobStuplb_02}):
\begin{equation}\label{2ra}
\begin{array}{c}
n_e= \displaystyle{n_{0} \over {\sqrt{1-v_{e}^2}}}\bigg\{ \big(
1-\left(1+4\sin^2 \theta _W\right)^{-1}{{\vec {\zeta}}_{e}}{{\vec
{v}}}_e\big ) \big( 1-{{\vec{\beta}}}{\vec{v}}_e \big)-\\-
\displaystyle\left(1+4\sin^2 \theta _W\right)^{-1}\sqrt{1-v^2_e}
\bigg[ { ({\vec{ \zeta}}_{e}{\vec{v}}_e)({\vec{\beta}}{\vec{v}}_e)
\over 1+\sqrt{1-v^2_e} }-{\vec{\zeta}}_{e}{\vec{\beta}} \bigg] +
O\big({1 \over \gamma}\big)
 \bigg\}.
\end{array}
\end{equation}

Using Eqs. (\ref{ra11}) and (\ref{0ra}) for the solid angle
distribution we get
\begin{equation}\label{1ra}
\frac{dI_{G}}{d\Omega}=\frac{3}{8\pi}I_{G}
 \bigg\{\!\gamma^{-4}(1-\beta\cos\vartheta)^{-3}\!
-\gamma^{-6}(1-\beta\cos\vartheta)^{-4}\!+
\frac{1}{2}\,\gamma^{-8}(1-\beta\cos\vartheta)^{-5}\!\bigg\}.
\end{equation}
From the last formula it follows that the $SL\nu$ is strongly
beamed and is confined within the cone given by $\delta \theta
\sim \gamma ^{-1}$. It should be noted here that this is a common
feature of different contributions to the neutrino spin light
radiation which is the inherent property of radiation by
ultra-relativistic particles.

In conclusion, we have discovered a new mechanism of the
electromagnetic radiation emitted by a neutrino with non-vanishing
magnetic moment  moving in the background matter and external
electromagnetic field ("spin light of neutrino"). The
generalization to the case of a neutrino with non-zero electric
dipole moment is just straightforward. Our general new result is
the prediction of a new type of electromagnetic radiation that is
emitted by a massive neutral particle with non-vanishing magnetic
(and/or electric) moment moving in the background matter. The
$SL\nu$ originates from the quantum spin flip transitions.
Therefore, the initially unpolarized neutrino beam can be
converted to the totally polarized beam composed of only
right-handed neutrinos $\nu_{R}$ in the considered radiation
process if the right-handed neutrinos are sterile states, i.e. do
not interact with the background environment.

We have also developed the quasi-classical theory of the $SL\nu$
that is valid in the case when the neutrino recoil can be
neglected or when the energy of the radiated photon is less than
the neutrino energy.

The considered radiation must be important for environments with
high effective densities, $n$, because the total radiation power
is proportional to $n^{4}$.  It is also shown that the $SL\nu$ is
strongly beamed in the direction of neutrino propagation and is
confined within a small cone given by
$\delta\theta\sim\gamma^{-1}$. The total power of the $SL\nu$  is
increasing with the neutrino energy increase and is proportional
to the fourth power of the neutrino Lorentz factor, $I \sim
{\gamma }^{4}$. It is also possible to show that the average
energy of photons of the spin light in matter is
\begin{equation}
\omega_{\gamma} \sim G_{F} n_{e} \gamma^{2}.
\end{equation}
Thus, the spin light emission rate in this case is proportional to
$\gamma^2$,
\begin{equation}\label{W}
\Gamma_{SL}={\sqrt{2} \over 3}\gamma^{2}\mu^{2}G_{F}^{3}n^{3}.
\end{equation}
From these estimations we predict that the $SL\nu$ is effectively
produced if the neutrino energy and matter density are large. Such
a situation can be realized in the dense matter of the early
Universe.

It is interesting to compare the rate of the $SL\nu$ in matter
with the rate of the Cherenkov radiation in magnetic field that is
widely discussed in literature (see, for example,
\cite{IoaRaf97}). First of all, in contrast to the Cherenkov and
transition radiations by neutrino, the spin light is produced by
neutrino even in the case when the refractive index of photons in
the background matter is equal to $n^{ref}_{\gamma}=1$. Let us
compare the rate of the neutrino spin light, Eq.(\ref{W}), with
the rate of the Cherenkov radiation in magnetic field given by
formula (25) of Ref.\cite{IoaRaf97}. If the magnetic field is less
than the electron critical field, $B<B_{0}=4.41\times 10^{13} \
G$, the ratio of the two rates is
\begin{equation}\label{ratio}
{\Gamma_{SL}\over \Gamma_{Ch}}=3.4\times
10^{8}{\gamma^{2}\mu^{2}G_{F}n^{3}\over p_{0}^{5}}\Big({B_{0}\over
B}\Big)^{6},
\end{equation}
where $p_{0}$ is the neutrino energy. This ratio is equal to
\begin{equation}\label{ratio_1}
{\Gamma_{SL}\over \Gamma_{Ch}}=13
\end{equation} for the two sets of
neutrino energy, strength of magnetic field and density of matter:
\begin{description}
  \item[1)] $p_0=10\ MeV, \ B=10^{-4}B_{0}, \ n=10^{30} cm^{-3},$
  \item[2)] $p_0=1\ MeV, \ B=10^{-3}B_{0}, \ n=10^{31} cm^{-3}$.
\end{description}
In both cases we take the neutrino mass equal to $m_{\nu}=1 \ eV$
and magnetic moment equal to $\mu=0.3\times 10^{-10}\mu_{0}$ that
is near the present experimental limit for the electron neutrino.
The neutrino spin light rate also dominates over the Cherenkov
radiation rate in strong magnetic field ($B\geqslant B_{0}$) if
the matter density is increased to the level of $n\geqslant
10^{33} cm^{-3}$. However, for such densities, neutrino magnetic
moment and energies in the range span from $1\ MeV$ to $10\ MeV$
the quantum approach \cite{Stu_03} to the $SL\nu$ has to be used.
If we consider the case of much higher density, $n=10^{37}
cm^{-3}$, and smaller magnetic moment,
 $\mu=0.3\times 10^{-16}\mu_{0}$,
and neutrino energy $p_0=1\ MeV$ then the quasi-classical approach
to the spin light is valid and the ratio of the two rates is given
again by Eq.(\ref{ratio_1}). Thus, we predict that the spin light
of neutrino can be more effective than the neutrino Cherenkov
radiation if the density of matter is rather large.

The spin light of neutrino together with the synchrotron mechanism
of radiation by charged particles, should be important for
understanding of astrophysical phenomena where powerful beams of
gamma-rays are produced.  One of the possible examples could be
the gamma-ray bursts.

We would like to thank Leo Stodolsky for helpful discussions.

\end{document}